\documentclass[12pt]{article}

\usepackage{times}
\usepackage{epsfig,amsmath}
\usepackage{subfigure}
\usepackage{graphicx}
\usepackage{color}
\usepackage{epstopdf}
\usepackage{cite}

\newenvironment{sciabstract}{%
\begin{quote} \bf}
{\end{quote}}

\newcounter{lastnote}

\title{Experimental Collision-Free Dominant Boson Sampling}

\author
{Jun Gao,$^{1,2,3,\dagger}$ Zhi-Qiang Jiao,$^{1,3,\dagger}$ Ruo-Jing Ren,$^{1,3}$ Xiao-Wei Wang,$^{1,3}$\\
Xiao-Yun Xu,$^{1,3}$ Wen-Hao Zhou,$^{1,3}$, Lu-Feng Qiao,$^{1,3}$ Xian-Min Jin$^{1,2,3,\ast}$\\
\\
\normalsize{$^1$School of Physics and Astronomy, Shanghai Jiao Tong University, Shanghai 200240, China}\\
\normalsize{$^2$Institute for Quantum Science and Engineering and Department of Physics,}\\
\normalsize{Southern University of Science and Technology, Shenzhen 518055, China}\\
\normalsize{$^3$Synergetic Innovation Center of Quantum Information and Quantum Physics,}\\
\normalsize{University of Science and Technology of China, Hefei, Anhui 230026, China}\\
\normalsize{$^\dagger$These authors contributed equally to this work}\\
\normalsize{$^\ast$E-mail: xianmin.jin@sjtu.edu.cn}\\
}

\date{}

\begin{document}
\baselineskip24pt

\maketitle

\begin{sciabstract}

Quantum computation, aiming at tackling hard problems beyond classical approaches, has been flourishing with each passing day. Unfortunately, a fully scalable and fault-tolerant universal quantum computer remains challenging based on the current technology. Boson sampling, first proposed by Aaronson and Arkhipov, is commonly believed as the most promising candidate to reach the intermediate quantum computational milestone, namely, quantum supremacy. Following this leading proposal, many experimental implementations as well as variants of boson sampling have been shown. However, most of these works are limited to small scale and cannot fulfill the permanent-of-Gaussians conjecture. Here, we experimentally demonstrate the largest scale boson sampling in the collision-free dominant regime using multi-port interferometer in a 3D photonic chip. We measure all 6,545 no-collision output combinations and validate the experimental results. Our work shows the potential of 3D photonic chip platform and represents a solid step toward large scale boson sampling.

\end{sciabstract}

Over the past few decades, quantum computation has become an emerging field due to its capability of finding ways to speed up the solution of classically hard computational problems\cite{QC}. Along with theoretical advances of different quantum algorithms\cite{Shor,Grover}, many physical systems have been experimentally investigated for encoding a qubit and processing various quantum tasks\cite{QC}. Despite these enormous advances, the construction of a universal quantum computer which can simultaneously meet all the technical demands\cite{DP2001} is considered as a long-term challenging goal. An intermediate but decisive milestone, quantum supremacy\cite{Supremacy}, could be achieved when an unequivocal quantum machine can solve a computational problem beyond the capability of any existed classical computer. To reach this goal, Aaronson and Arkhipov formalized a well-defined but also experimentally friendly scheme called the boson sampling problem\cite{AA}, as the near team attainable demonstration of quantum supremacy\cite{Supremacy}.

The main idea of boson sampling is to estimate the output distribution of $n$ indistinguishable photons scattering in an $m$-mode interferometer described by a Haar random matrix $U$. Based on two unproven but reasonable conjectures, Aaronson and Arkhipov showed that for a classical computer, it is hard to give either the exact or approximate results of the scattering amplitude. The computational complexity of boson sampling problem mainly rooted in calculating the submatrix permanent of $U$. This argument holds only when $m$ is large enough (at least much larger than $n^{2}$), such that the submatrices of $U$ can be regarded as approximately Gaussian, which is the requirement of permanent-of-Gaussians conjecture. Another cause of the complexity behind boson sampling is that there should be $\left(\begin{array}{l}{m} \\ {n}\end{array}\right)$ output combinations in total, and this factor renders the estimation of the output distributions even more complex\cite{Complex}.

Compared to other physical systems\cite{Duan}, photon seems to offer the most suitable platform from the experimental perspective. Photonic scheme only requires single photon sources, linear optics network and on-off photodetection, which are all accessible within the current technology. Ever since Aaronson and Arkhipov proposed such scheme, there have been several experimental implementations of boson sampling\cite{Oxford_2013,Vienna&Jena_2013,Roma&Milano_2013,Roma&Milano_Birthday_2013,Brisbane_2013,Roma&Milano_validation_2014,Bristol_Haar_random_2014,Bristol_Universal_2015,Vienna&Jena_2015,Roma&Milano_2016,Brisbane_QD_2017,USTC_time_bin_2017,USTC_QD_2017,USTC_Lossy_2018} as well as the variants of boson sampling, like scattershot boson sampling\cite{Roma&Milano_Scattershot_2015,USTC_Scattershot_2018} and Gaussian boson sampling\cite{Bristol_Gaussian_2019}. However, compared to the photon number $n$, the scale of the interferometers $m$ of most experiments cannot follow the original conjecture since boson sampling is supposed hard only in this regime.

Here, we report experimental large scale boson sampling using a 35-mode 3D femtosecond laser writing photonic chip with 3-photon injection. The full dimension of this unitary transformation in Hilbert space is 7,770, which is by far the largest scale in the integrated photonics platform. With self-developed multi-channel coincidence module (MCCM, see Methods for device details), we simultaneously measure all 6,545 collision-free output events and further validate the results against uniform and distinguishable samplers. These results shows the capability of 3D structure and femtosecond laser writing technique. Such features combining on-chip photon generation\cite{SFWM} and imaging based large scale quantum correlation measurement\cite{Cospli} could open up a new regime for boson sampling.

\begin{figure}[htbp]
	\centering
	\includegraphics[width=1.0\linewidth]{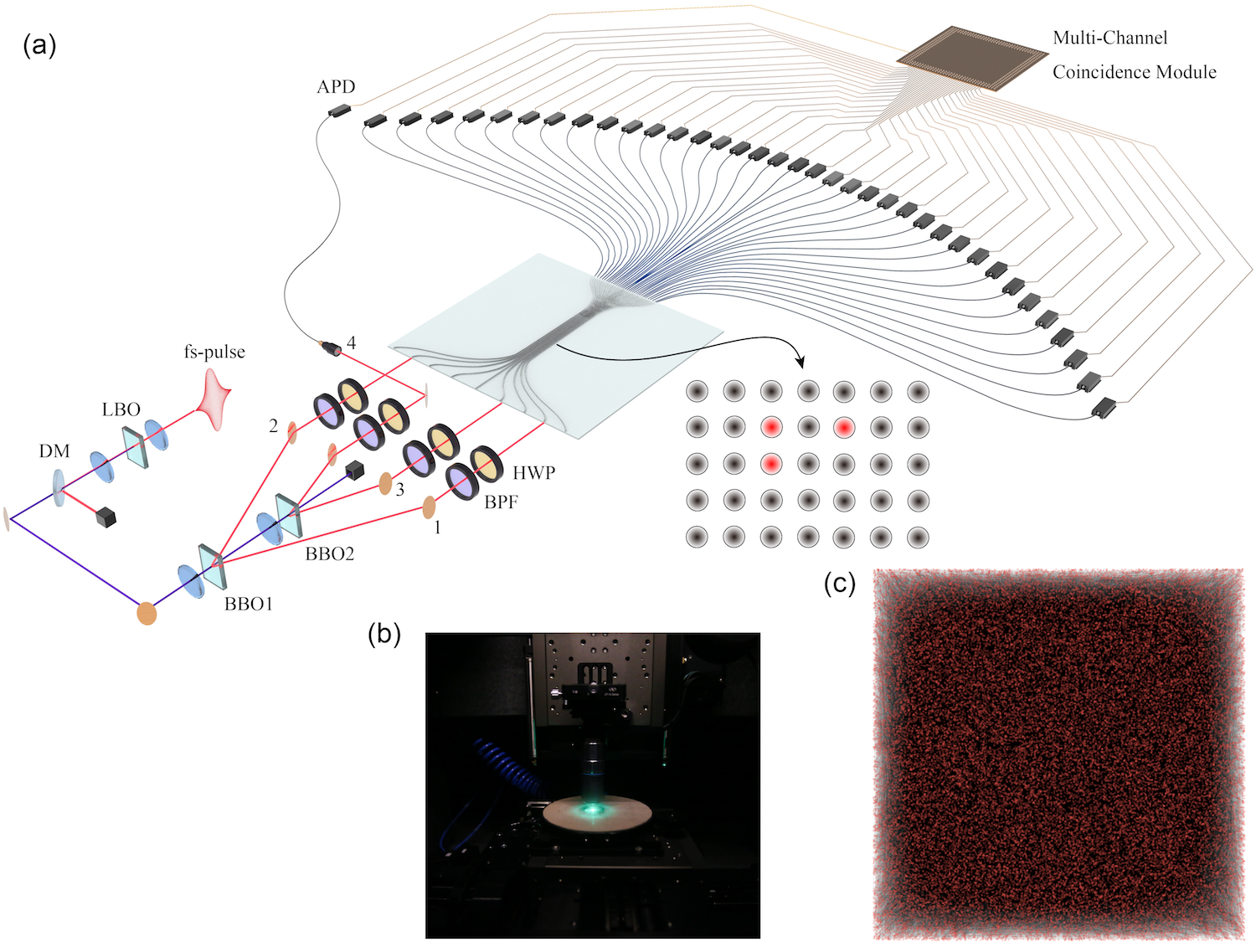}
	\caption{\textbf{Schematic of boson sampling experiment.} \textbf{(a)} The multi-photon fock states are prepared by successively pumping two BBO crystals with a frequency doubled 390nm fs laser pulse. The generated non-collinear down converted photon pairs are collected by four fiber couples and three of them (1,2,3) are guided to the photonic chip through a PM V-groove fiber array while the last photon (4) serves as the trigger. All the photons are filtered by BPF and polarization compensated by HWP to guarantee the indistinguishability. Two linear motorized translation stages are used to scan the temporal delay among these injected photons. After leaving the photonic chip, all 35 modes are connected to a large array of APD. The electronic signals are all processed by our home-made MCCM, which can collect all collision-free events in the large Hilbert space simultaneously. The cross section of the interference zone in the photonic chip. The two dimensional structure allows the photonic chip to implement a Haar random matrix exponentially fast. The red spots represent the injection ports. \textbf{(b)} Femtosecond laser writing setup. \textbf{(c)} The hypercube graph describes the 3-photon injection equivalent full Hilbert space dimension.}
	\label{f1}
\end{figure}

We prepare the 3D photonic chip using femtosecond laser direct writing technique\cite{Fab} (see Methods for fabrication details), as shown in Fig.\ref{f1}(b). A high power femtosecond pulse (290-fs centered at 513nm) is focused inside the borosilicate glass substrate to introduce a permanent refractive index change. Translating the sample with respect to the laser allows us to fabricate 3D structure in the substrate. The propagation direction is mapped to time evolution, and the 2D cross section structure has been proven versatile for many quantum tasks\cite{QW_2D,FH}. The central interference zone of the chip is made of 35 waveguides with a 7$\times$5 2D cross section. The 2D structure transforms to a planar configuration at both input and output sides to match the one dimensional fiber array with 8 input and 35 output modes. We randomly vary the S-bend over the transforming zones at both sides to introduce random phase shift\cite{Roma&Milano_2013} and coupling length, theoretical work has shown that such structure renders the unitary transformation converge to Haar random matrix exponentially fast\cite{Haar}, which can also decrease the total loss of the photonic chip. The basic connections between the nearest sites of the cross section can be described by a 2D grid graph. By injecting multi-photon states, the grid graph will extend to a high dimensional hypercube graph. We show the 3-photon equivalent hypercube graph in Fig.\ref{f1}(c) derived from the Cartesian product of three grid graphs, which shows very complex connectivity between different modes. The node number of the hypercube graph is 42,875, for boson sampling problem, due to the photons’ symmetry, the full dimension of state space will become 7,770. 

The experimental setup is schematically shown in Fig.\ref{f1}(a). A femtosecond pulse (140 fs) with a central wavelength of 780nm from a modelocked Ti:sapphire laser is first frequency-doubled in a lithium triborate (LBO) crystal. The generated 390nm laser pulse then successively passes through two beta barium borate (BBO) crystals and produces two pairs of correlated photons via spontaneous parametric down-conversion. The first BBO crystal is type-II phase matching in a beam-like scheme\cite{Beamlike} while the next BBO crystal is cut for type-I phase matching\cite{TypeI}. The generated photons are filtered by 3nm band pass filters (BPF) to achieve near-identical spectrum and then coupled into polarization-maintaining (PM) optical fibers with proper polarization control to ensure the same polarization of different modes. Photons are injected into the photonic chip via a butt-coupled PM V-groove fiber array. Temporal indistinguishability for non-classical interference among the down converted photons is achieved with outside delay lines. The output of the waveguides are coupled to multimode V-groove fiber array and then connected to the avalanche photodetectors (APD). All the electronic signals are sent to the MCCM, which can process all the photon scattering output combination events simultaneously.

\begin{figure}[htbp]
	\centering
	\includegraphics[width=1.0\linewidth]{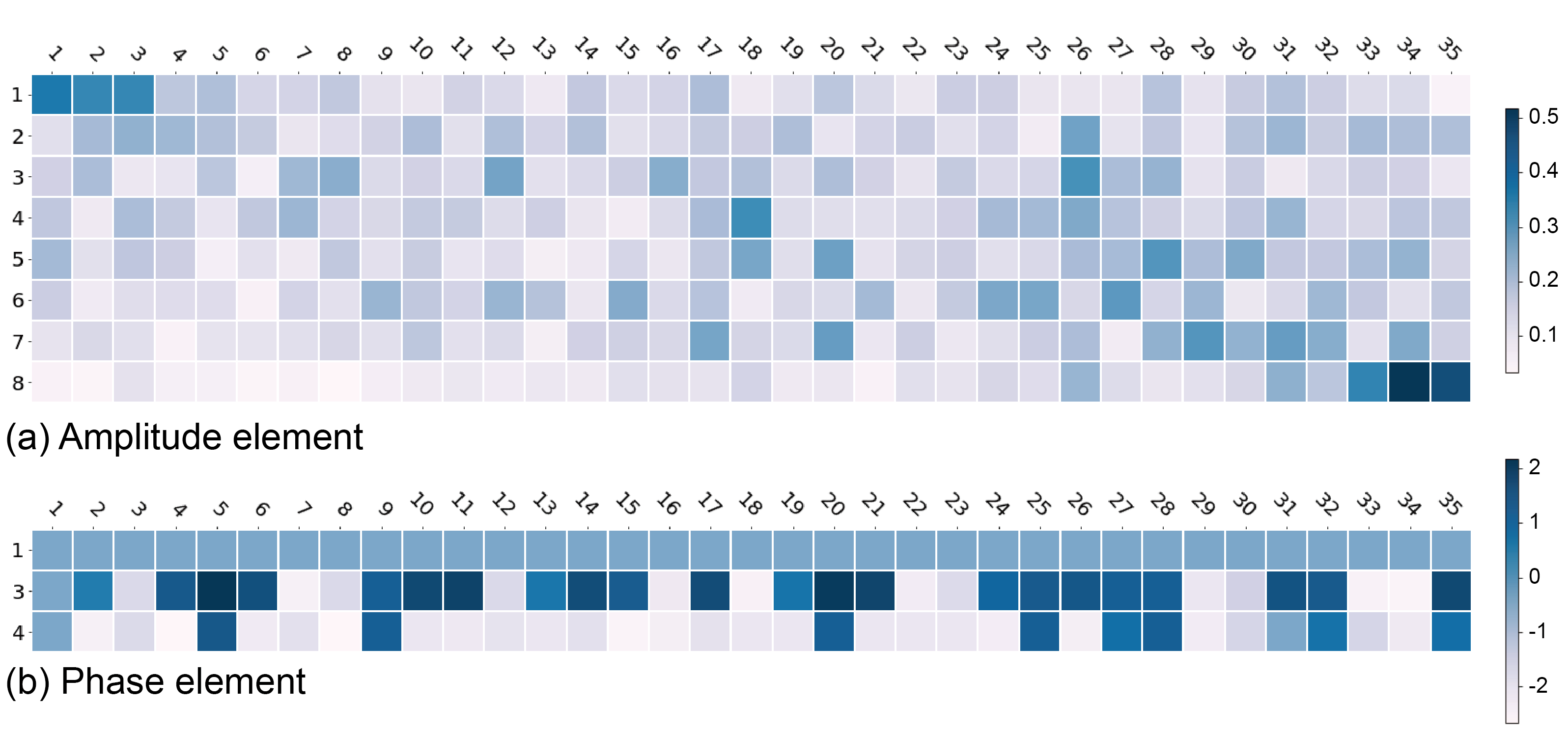}
	\caption{\textbf{Experimental characterization of the scattering matrix.} \textbf{(a)} The amplitudes of the scattering matrix with 8 inputs. These results are measure by heralded single photon injection. Here we choose port 1, 3 and 4 as the injection ports. \textbf{(b)} The relative phase between different selected injection ports. These values are reconstructed by scanning hundreds of HOM interference among these ports.}
	\label{f2}
\end{figure}

After all these preparation works, we need to characterize the unitary matrix describing the photonic chip (see Methods for further details). We first inject heralded single photons into the chip with different inputs and measure all 35 output modes simultaneously, from which we can obtain all the moduli of the scattering matrix. The experimentally measured amplitude is demonstrated in Fig.\ref{f2}(a), from which we choose port 1, 3 and 4 as the input ports considering both the transmission efficiency and the randomness. Then we perform a series of HOM interference among these ports to determine the relative phases between different modes\cite{Characterization}. The histogram of reconstructed phase is shown in Fig.\ref{f2}(b). After all these characterizations, we can get a $3\times35$ matrix which fully describes the scattering process in the experiment. 

\begin{figure}[htbp]
	\centering
	\includegraphics[width=0.95\linewidth]{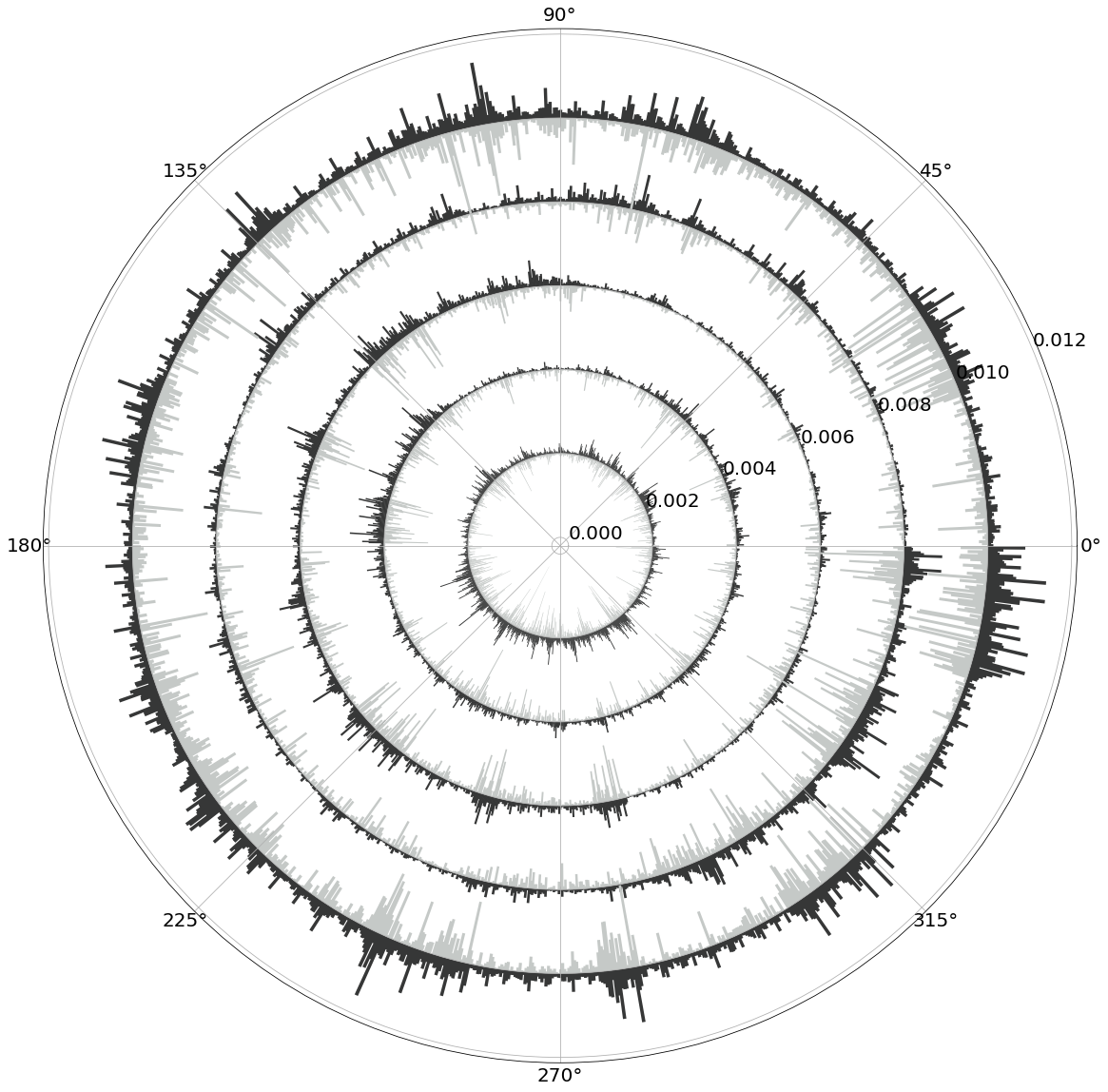}
	\caption{\textbf{Experimental and theoretically predicted distributions of all collision-free multi-photon events.} With 3 indistinguishable photon injection into 35 spatial modes, we measure 6,545 no-collision output combinations in total. After normalizing the distribution probability, we list all these results on a five-layer ring plot from output (1,2,3) to (33,34,35). The gap between different rings are 0.002. Experimental results are plotted in dark gray while the theoretical predictions are plotted in light gray. The fidelity $F=\sum_{i} \sqrt{s_{i} t_{i}}$ between these two distributions is 89.3\%.}
	\label{f3}
\end{figure}

We implement the boson sampling experiment in the photonic chip with photon 1, 2 and 3. Photon 4 is regarded as the trigger event which postselects the fourfold coincidence. We scan the delay lines successively to achieve the genuine three-photon interference in the interference section. The pump power is selected at a moderate level as 700mW . After a collection time of near 50 hours, we obtain 570,312 fourfold coincidence events in total. After normalizing the probability distribution, we plot the experimental results ($s_{i}$) (dark gray) of all 6,545 collision-free combinations from (1,2,3) to (33,34,35) in a five-layer ring, as shown in Fig.\ref{f3}, and compare the results with theoretical prediction ($t_{i}$) (light gray) from the reconstructed matrix. The experimental distribution has a fidelity $F=\sum_{i} \sqrt{s_{i} t_{i}}$ of 89.3\% with theoretical prediction. We also calculate the total variation distance, which is defined as $D=(1 / 2) \sum_{i}\left|s_{i}-t_{i}\right|$, and the value is 0.350. The main reason of this deviation from the theoretical prediction is that the second transforming zone after the interference section introduces differential losses for different channels, which further influences the out-coming results. This shortcoming can be overcome in the future by adopting the imaging-based large scale correlation measurement\cite{Cospli} since we could directly measure the 2D cross section structure.

\begin{figure}[htbp]
	\centering
	\includegraphics[width=1.0\linewidth]{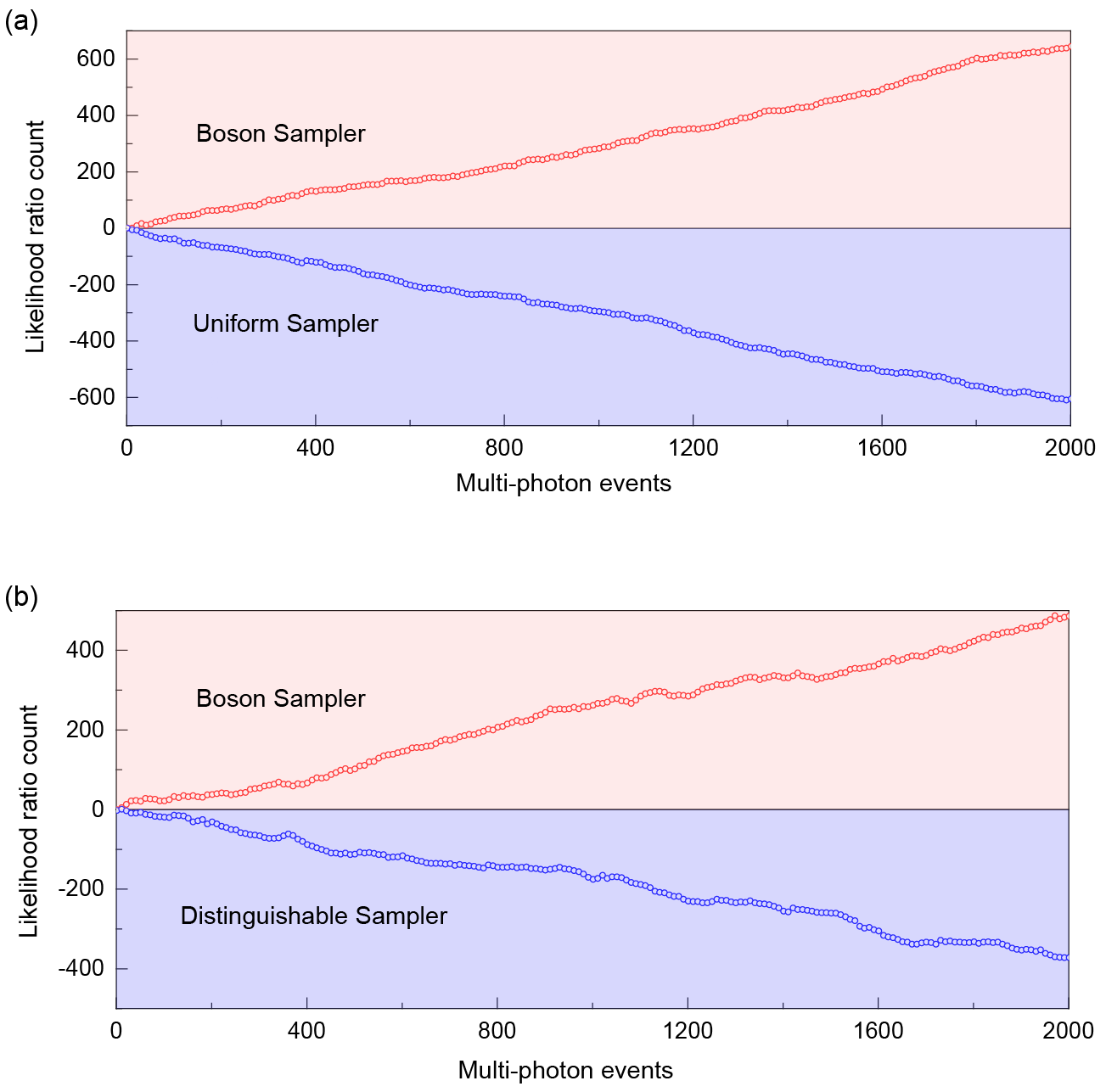}
	\caption{\textbf{Validation of experimental boson sampler against other classical samplers.} Validation our experimental results against uniform (distinguishable) sampler are plotted in the \textbf{a} (\textbf{b}) figure. Unlike any classical computing imposters, the results of a genuine boson sampler increases almost monotonically according to the calculation result of the discriminator. Only after a few hundreds events, our experimental results prominently differ from the classically simulated imposters.}
		\label{f4}
\end{figure}

Unlike NP problems, boson sampling is not efficiently verifiable due to its sampling task nature, especially for calculating the full probability distribution of a large scale boson sampling experiment. To validate that our experimental results against uniform sampler and distinguishable sampler, we perform both the row-norm estimator test\cite{Uniform} and the likelihood ratio test\cite{Roma&Milano_validation_2014}. We can introduce a certain discriminator and a counter (see Methods for further calculating details), the counter will update according to the calculation result of the discriminator. For a boson sampler, the counter increases almost monotonically. As shown in Fig.\ref{f4}(a) and (b), the red dots are based on experimental data while the blue dots are the test results from simulated data. We can infer from the difference that our experimental device is a genuine boson sampler even only after calculating a few hundreds multi-photon events, providing experimental support for both the efficiency and scalability of these methods, especially for future verification of classically intractable situations.

\begin{figure}[htbp]
	\centering
	\includegraphics[width=1.0\linewidth]{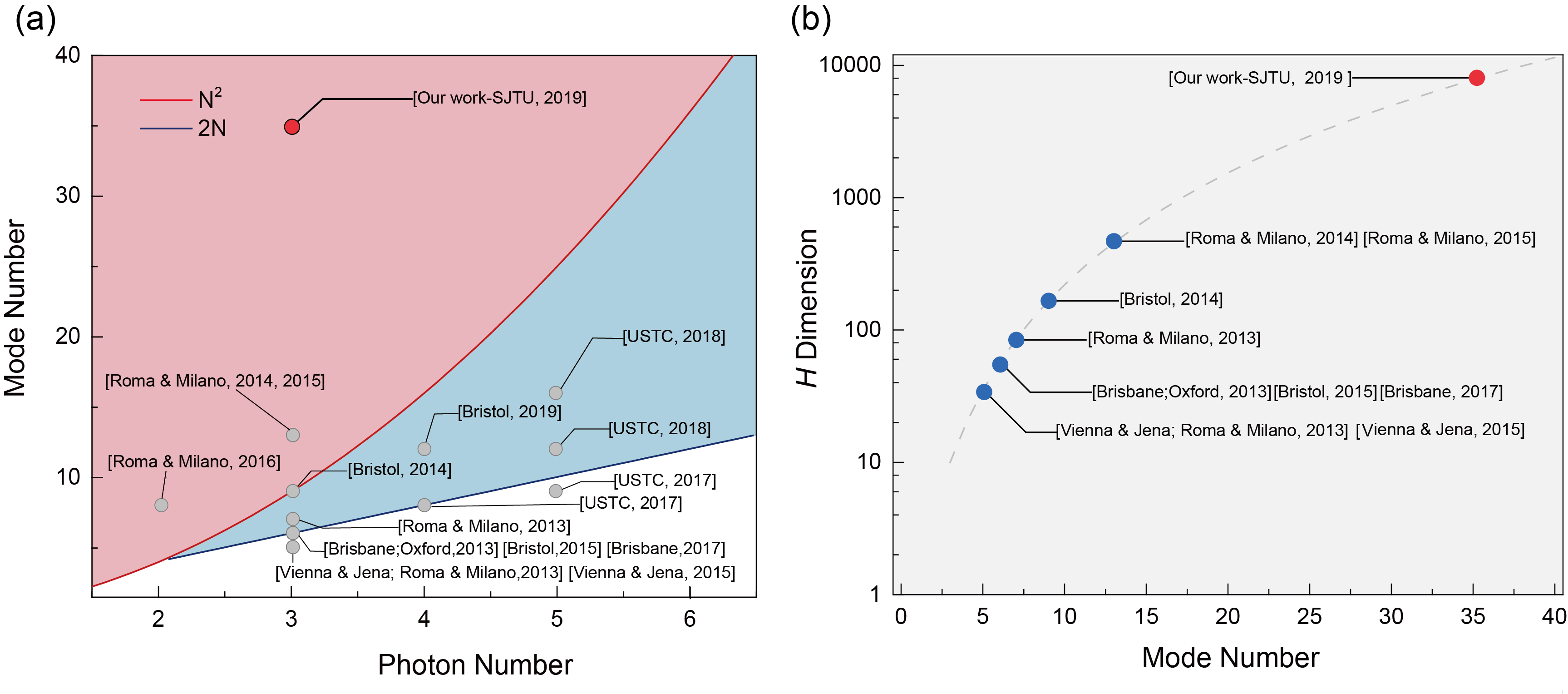}
	\caption{\textbf{Summary of the state-of-the-art boson sampling experiments.} \textbf{(a)} The photon number (N) and the mode number (M) of all current boson sampling experiments are listed in three different regimes according to the relation between M and N. Most of the previous experiments are missing in the regime over $m\gg n^{2}$. The mode number of our photonic chip is large enough to satisfy the permanent-of-Gaussians conjecture, thus allows our device working in the collision free dominant regime. \textbf{(b)} A comparison of Hilbert space dimension among all three photon experiments. As the mode number increases, the dimension of the Hilbert space grows exponentially, which further enhance the complexity of boson sampling problem.}
	\label{f5}
\end{figure}

We summarize the state-of-the-art boson sampling experiment in Fig.\ref{f4}(a). We list both the mode number of the unitary transformation and the photon number involved in the non-classical interference of different experiments. The graph is divided into three regimes according to the relation between the mode number and the photon number, which in related to the permanent-of-Gaussians conjecture. It is strongly suggested that at least under the condition of $m=O(n^{2})$, the unitary matrix is large enough to fulfill conjecture. Only in this regime, the small submatrix of the Haar random matrix looks approximately Gaussian, which doesn't have any special structure that can be utilized to cheat in classically simulating the distribution. Another interesting but harsh challenge is whether the hardness of approximate boson sampling still holds when $m=O(n)$, like $m=2n$, which unfortunately, remains an open question now. So the commonly suggested mode number should be at least larger than $m=n^{2}$. We can see in Fig.\ref{f5}(a) that most of the current experiments are still in the regime between $m=n^{2}$ and $m=2n$. Since the hardness of boson sampling problem is highly based on this unproven but plausible conjecture, the mode number of our work is sufficiently large, thus can fulfill the permanent-of-Gaussians conjecture for the first time. For further experimental investigation to realize quantum supremacy, it is essential that not only focus on expanding the physical scale, but also the theoretical assumption behind the problem. For future investigation

Another advantage of large mode number is that this factor could lead to a large Hilbert space dimension. The total complexity of simulating brute force boson sampling problem is to combine the all input–output possibilities and the time consumption in calculating each permanent. The increasing mode number will induce an explosive expand of the Hilbert space when fixing the photon number. As we show in Fig.\ref{f5}(b), we compare the total Hilbert space of all previous $n=3$ experiments with our work, the Hilbert space dimension is at least one order of magnitude larger than previous experiments. Even after Clifford and Clifford propose an algorithm to fully simplified the total complexity\cite{CC}, such mode number could be considered as useful resources in quantum simulation tasks like simulating molecular vibronic spectra and spin chain\cite{Spec,Spin}. A combination of large photon number and large spatial number together will surely enlarge the exploring boundary of quantum information processing.

For future boson sampling experiment, it is very important that we should seek other possible resources to surpass the capability of classical algorithm\cite{Classical}. Resources like polarization and time tin should be all put altogether into fock state encoding to reach the limitation of classical supercomputer. Luckily, there have been some pioneer trials on such degree of freedoms\cite{USTC_time_bin_2017,USTC_Scattershot_2018} independently. 

In conclusion, we have experimentally demonstrated the largest mode number boson sampling experiment in the quantum integrated photonic platform. The 3D structure allows our chip to realize the Haar unitary transformation in a more efficient way. The mode number of the photonic chip is sufficiently large that we can fulfill the permanent-of-Gaussians conjecture and measure the output results in the collision-free dominant regime. We further verify our experimental results are sampled from genuine boson sampler. Quantum integrated photonic platform has been proved versatile for both on-chip quantum state generation and quantum information processing\cite{PhotoReview}. Along with on-chip quantum light source, imaging-based large scale correlation measurement and lossy scattershot boson sampling scheme, our experimental approaches broaden the way towards quantum supremacy through boson sampling.\\

\subsection*{Acknowledgments.}
The authors thank Barry Sanders, Peter Rohde, Martin Plenio, Ish Dhand and Jian-Wei Pan for helpful discussions. This research is supported by National Key R\&D Program of China (2017YFA0303700), National Natural Science Foundation of China (NSFC) (61734005, 11761141014, 11690033), Science and Technology Commission of Shanghai Municipality (STCSM) (15QA1402200, 16JC1400405, 17JC1400403), Shanghai Municipal Education Commission (16SG09, 2017-01-07-00-02-E00049), X.-M.J. acknowledges support from the National Young 1000 Talents Plan.\\

\subsection*{Methods}
\paragraph*{Fabrication of the 3D boson sampling chip:} The 1D 9-port fan-in, 35-port fan-out and 2D $7\times5$ random coupling structure is fabricated in a borosilicate glass substrate. We feed a 513nm femtosecond laser (upconverted from a pump laser of 1,026nm, 290 fs pulse duration, 1MHz repetition rate) into a cylindrical lens to remake the laser beam shaping into a long and narrow one. We then focus the pulse onto the borosilicate substrate with a $\times50$ objective lens (numerical aperture of 0.55) at a constant velocity of 15mm/s. And the 3D contour is created by the permanent laser-induced refractive index increase. The matrix randomness is introduced by different coupling length and varied coupling separation distance among these ports. Power compensation is used to improve uniformity of waveguides.

\paragraph*{Home-made multi-channel coincidence module (MCCM):} Our self-developed multi-channel coincidence module (MCCM), has forty import channels with independently adjustable delay time. MCCM can record all the single and multifold coincidence events at the same time. It records all the electronic signals from 35 avalanche photodetectors and processes all 6,545 photon scattering output combination events simultaneously. Single collection time can be set up to hours and recycle times can be more than 1,000. In our experiment, we collect the data of near 50 hours, and obtain 570,312 fourfold coincidence events in total.

\paragraph*{Characterization of the boson sampling chip:} The boson sampling photonic chip can be described by a unitary matrix, of which each element is a complex number. The characterization of this matrix can be completed by two steps. The first one is to determine the moduli of each element. This step can be easily achieved by injecting coherent state or heralded single photon and measuring the output intensity distribution. The next step is to retrieve the arguments of each element. The absolute value of the arguments is first obtained by two photon HOM interferences. To further determine the signs of the arguments, extra interferences between different channels are scanned, thus we can reconstruct the full information of the unitary matrix.

\paragraph*{Validation of boson sampling data:} To verify the experimental data are extracted from a genuine boson sampler other than classical computing imposters. The most common ones are uniform sampler and distinguishable sampler. To distinguish from a uniform sampler, Aaronson and Arkhipov propose row-norm estimator (RNE) test. For each event $k$, we can introduce an estimator $P_{k}=\prod_{i=1}^{n} \sum_{j=1}^{n}\left|A_{i, j}\right|^{2}$, here $A$ is the submatrix related to the events. Then we can update a counter following the rules,
\begin{equation}
C=\left\{\begin{array}{l}{C+1, \text { if } P_{k}>(n / m)^{n}} \\ {C-1, \text { else }}\end{array}\right.
\end{equation}
For a boson (uniform) sampler, the counter will be above (below) zero even after a few events.
Next, we use likelihood ratio test to exclude the possibilities of distinguishable samplers. Like RNE test we can also calculate an estimator $L_{k}=p_{k}^{i n d} / q_{k}^{d i s}$. Here $p_{k}^{i n d}$ and $q_{k}^{d i s}$ are the corresponding probabilities related to the event from different kinds of samplers. We use the following rules to update the counter $C=0$,
\begin{equation}
C=\left\{\begin{array}{lr}{C,} & {a_{1}<L_{k}<1 / a_{1}} \\ {C+1,} & {1 / a_{1} \leq L_{k}<a_{2}} \\ {C+2,} & {L_{k} \geq a_{2}} \\ {C-1,} & {1 / a_{2} \leq L_{k}<a_{1}} \\ {C-2,} & {L_{k} \leq 1 / a_{2}}\end{array}\right.
\end{equation}
In our experiment, we choose $a_{1}=0.9$ and $a_{2}=1.5$. Again, after a few events, for a genuine boson sampler, $C$ increases almost monotonically.
\clearpage

\clearpage

\clearpage

\clearpage

\clearpage

\clearpage

\clearpage

\end{document}